# Impact Of A Uniform Plasma Resistivity In MHD Modelling Of Helical Solutions For The Reversed Field Pinch Dynamo

D. Bonfiglio[a], S. Cappello[a] and D. F. Escande[a,b]

[a]*Consorzio RFX, Associazione EURATOM-Enea sulla fusione, Padova, Italy*
[b]*CNRS-Université de Provence, Marseille, France*

**Abstract.** Till now the magnetohydrodynamic (MHD) simulation of the reversed field pinch (RFP) has been performed by assuming axis-symmetric radial time independent dissipation profiles. In helical states this assumption is not correct since these dissipations should be flux functions, and should exhibit a helical symmetry as well. Therefore more correct simulations should incorporate self-consistent dissipation profiles. As a first step in this direction, the case of uniform dissipation profiles was considered by using the 3D nonlinear visco-resistive MHD code SpeCyl. It is found that a flattening of the resistivity profile results in the reduction of the dynamo action, which brings to marginally-reversed or even non-reversed equilibrium solutions. The physical origin of this result is discussed in relation to the electrostatic drift explanation of the RFP dynamo. This sets constraints on the functional choice of dissipations in future self-consistent simulations.

## I. INTRODUCTION

The reversed field pinch (RFP) is a toroidal device for plasma magnetic confinement. Its magnetic configuration is akin to the tokamak, but the amplitude of the poloidal field is comparable to the toroidal one, which reverses in the outer region. A self-organized plasma flow provides, through the Lorentz force, the poloidal component of the electromotive force necessary to sustain the highly sheared configuration: this is the so-called RFP dynamo action. During the last years, the increasing evidence in different RFP experiments (see Ref. 1 and references therein) of a tendency to develop regimes characterized by a good degree of helical symmetry has motivated renewed interest in the modelling of the RFP as a saturated regime of a single kink-type magnetohydrodynamic (MHD) mode. Three-dimensional visco-resistive MHD numerical simulations of the RFP display a transition from turbulent states to helical laminar equilibria when dissipation is increased[2,3]. Till now, the RFP has been simulated by assuming time-independent dissipation coefficients with axis-symmetric radial profiles. In the case of helical states[2,3,4,5], this assumption is not justified since these dissipations should be constant on helical magnetic flux surfaces. Therefore, more correct simulations should incorporate self-consistent dissipation profiles. As a first step in this direction, the effect of uniform dissipation profiles in MHD modelling of helical RFP solutions is considered in this paper.

## II. NUMERICAL RESULTS

The RFP is simulated with the SpeCyl 3D non-linear visco-resistive MHD cylindrical code[6]. SpeCyl provides numerical solutions of the compressible non-linear MHD model in the constant-pressure constant-density approximation, including resistivity, $\eta$, and viscosity, $\nu$, whose equations in dimensionless units are

$$\frac{\partial \mathbf{B}}{\partial t} = -\nabla \times \mathbf{E} = \nabla \times (\mathbf{v} \times \mathbf{B} - \eta \mathbf{J}), \quad \frac{\partial \mathbf{v}}{\partial t} + (\mathbf{v} \cdot \nabla)\mathbf{v} = \mathbf{J} \times \mathbf{B} + \nu \nabla^2 \mathbf{v}, \quad (1)$$

where $\mathbf{J} = \nabla \times \mathbf{B}$ and $\nabla \cdot \mathbf{B} = 0$. An alternative form of these equations can be obtained, in which the two dimensionless parameters $(H, P)$ are involved, namely the Hartmann number $H \equiv (\eta \nu)^{-1/2}$ and the magnetic Prandtl number $P \equiv \nu/\eta$. This highlighted the leading role of $H$ in the MHD dynamics of RFP numerical simulations[2,3].

To study stationary sustainment of helical solutions, Equations 1 are numerically solved in the case of a single-helicity spectrum composed by the $m=1, n=-10$ mode and 79 of its poloidal harmonics. Plasma current and toroidal magnetic flux are constant, and thus the pinch parameter, $\Theta \equiv B_\vartheta(a)/\langle B_z \rangle = 1.6$. Resistivity and viscosity are used with profiles given by $\eta(r) = \eta_0 \left(1 + \alpha_\eta r^{\beta_\eta}\right)$ and $\nu(r) = \nu_0 \left(1 + \alpha_\nu r^{\beta_\nu}\right)$. Different values of the dissipative coefficients $\eta_0$ and $\nu_0$ are used, corresponding to Lundquist numbers $S \equiv \eta_0^{-1}$ in the range $1.5 \times 10^4 \leq S \leq 3 \times 10^5$ and magnetic Prandtl number in the range $100 \leq P \leq 3000$. For each choice of the couple $(S, P)$, two types of dissipation profiles are used. Profiles of the first type are characterized by $\alpha_\eta = \alpha_\nu = 19$ and $\beta_\eta = \beta_\nu = 10$, which implies that $\eta$ and $\nu$ are practically constant for $r \leq 0.8$, rising steeply in the edge region (this is the standard shape of dissipation profiles used in SpeCyl). The second case consists in uniform profiles $(\alpha_\eta = \alpha_\nu = 0)$.

The numerical evolution of reversal parameter $F \equiv B_z(a)/\langle B_z \rangle$ and volume averaged energy of magnetic fluctuations $\langle \delta E^M \rangle$ is shown in Figure 1. A non-reversed axis-symmetric ohmic state is used as the initial configuration for each simulation, at $t = 1000\, \tau_A$. A small perturbation, corresponding to the unstable $m=1, n=-10$ mode, is then applied to the system. The initial phase of MHD dynamics is qualitatively the same for standard profile (Fig. 1a) and uniform profile simulations (Fig. 1b): the mode amplitude grows exponentially, causing $F$ to change its sign, and non-linearly saturates. In the case of standard profile simulations, a helical RFP equilibrium is reached just after saturation. The following slow fall of $F$ is caused by the loss of toroidal flux due to numerical discretization effects ($\Theta = 1.609$ equilibria are marked with a cross in Fig. 1a). In the case of uniform dissipation, i.e. flat profiles simulations, the amplitude of the helical deformation decreases after saturation, leading to a reduction of the dynamo action. Consequently, the final helical equilibrium turns out to be marginally-reversed or even non-reversed in the set of simulations with pinch parameter $\Theta = 1.6$ (simulations at higher pinch parameter, $\Theta = 1.9$, show deeply-reversed equilibria also with uniform dissipation profiles).

The dependence of the amplitude of the equilibrium helical deformation on the Hartmann number is shown in Figure 2. In the case of standard profile simulations, the reversal becomes deeper with increasing $H$ and saturates at $F \cong -0.3$ for large $H$. Similarly, the magnetic perturbation energy grows and saturates at 5% of the total magnetic energy. The saturation of $F$ and $\langle \delta E^M \rangle$ is found as expected for single-helicity dynamics[6,7,8]. Saturation for large $H$ is also shown for uniform profile simulations, but in this case the magnetic fluctuation energy is one order of magnitude smaller than the one in the standard case, and the resulting high-$H$ equilibrium is only marginally reversed.

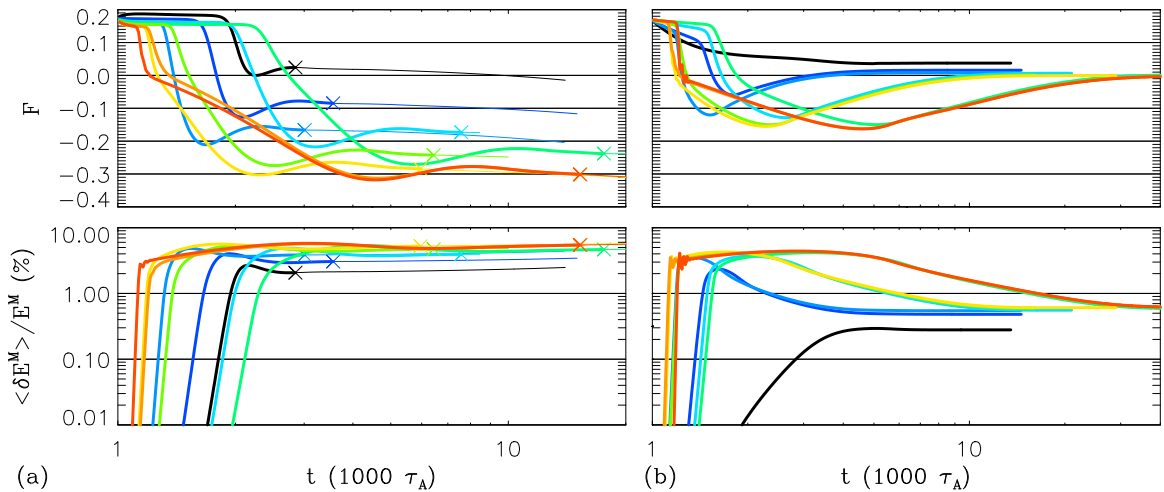

**FIGURE 1.** Temporal evolution of reversal parameter F and energy of the magnetic fluctuations for (a) standard profile simulations and (b) uniform profile simulations.

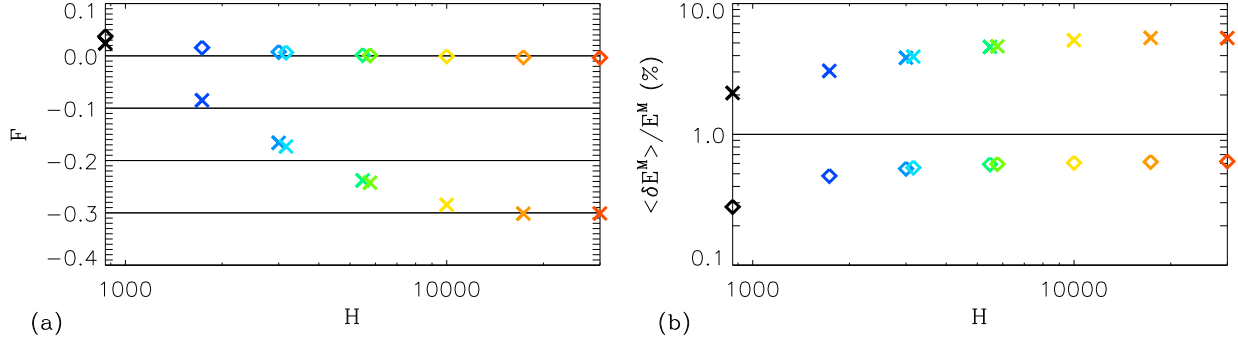

**FIGURE 2.** Dependence on the Hartmann number H of (a) reversal parameter F and (b) energy of the magnetic fluctuations. Crosses represent standard profile simulations, diamonds uniform profile simulations. Refer to Fig. 1 for the color code.

## III. DISCUSSION

The numerical results we have just shown, indicate helical equilibrium solutions to be strongly sensitive on the profiles of dissipation coefficients. To discriminate between effects due to resistivity and viscosity, a couple of simulations (not shown here) has been performed, one with uniform ν and standard η profile, the other with uniform η and standard ν profile. The first simulation is qualitatively similar to standard reversed simulations, while the second one turns out to be non-reversed. We therefore conclude that the decrease in magnetic fluctuations is due to the flattening of the resistivity profile, as also observed for multiple helicity simulations in Ref. 9.

The reason why a flattening of the resistivity profile results in a reduction of the dynamo action, can be explained in relation to the fact that the dynamo velocity field is mainly provided by an electrostatic drift, due to a spacially modulated electrostatic field which arises in the plasma as soon as the cylindrical symmetry is broken[10,11]. For a stationary equilibrium, the component of Ohm's law parallel to the magnetic field gives:

$$\eta\mu \equiv \eta \mathbf{J} \cdot \mathbf{B}/B^2 = \mathbf{E}_0 \cdot \mathbf{B}/B^2 - \nabla\phi \cdot \mathbf{B}/B^2, \qquad (2)$$

where $\mathbf{E}_0 \equiv E_0 \hat{\mathbf{z}}$ is the uniform toroidal induction electric field, and $-\nabla\phi$ is the spatially modulated electrostatic field. If the configuration is force-free, $\mathbf{J} \times \mathbf{B} = 0$, the parallel current density $\mu$ is a function of the helical flux function $\psi \equiv mRA_z + nrA_\vartheta$, where $\mathbf{A}$ is the vector potential. This is practically the case for these numerical simulations, which are nearly force-free. If the resistivity is also constant on magnetic surfaces, as in the trivial case of uniform profiles, the r.h.s. of Eq. 2 is function of $\psi$. A finite electrostatic field is hence needed in order to account for the helical modulation of $E_0 B_z/B^2$ along field lines. This is shown in Figure 3, where contributions to parallel Ohm's law are displayed for the (non-reversed) equilibrium of a uniform profile simulation.

In the case of standard profile simulations, $\eta\mu$ is not function of $\psi$, due to the radially-increasing resistivity profile. This situation is shown in Figure 4 for the equilibrium of a standard profile simulation[10]. When the field line passes by the plasma edge, $\eta\mu$ increases while $E_0 B_z/B^2$ becomes smaller. Therefore an electrostatic field, larger than the one needed in the case of uniform resistivity, is required in order to account for the helical modulation of $E_0 B_z/B^2 - \eta\mu$. The electrostatic drift due to this larger electrostatic field provides an enhanced dynamo action, which is sufficient to sustain the reversed configuration. We note that the presence of resistivity gradients along field lines has also been identified as the essential feature for the formation of a stationary *m*=1 convection cell in numerical MHD simulations of sawtooth activity in tokamaks[12].

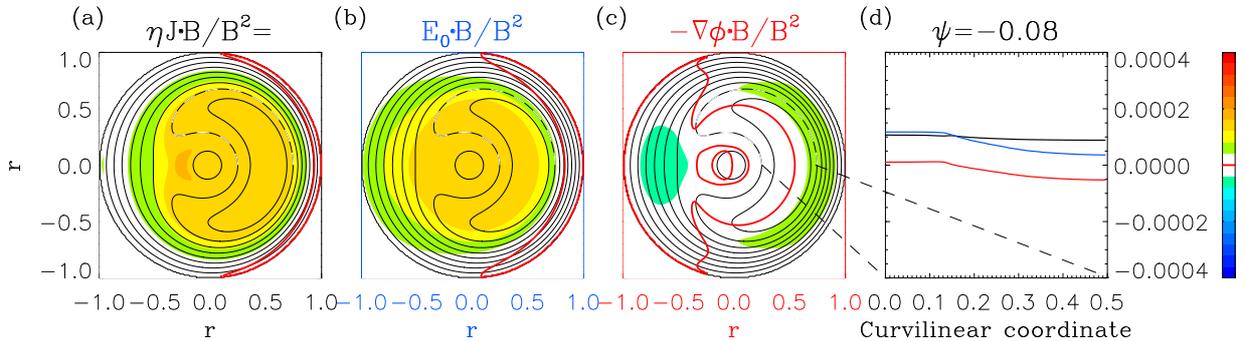

**FIGURE 3.** Contributions to parallel Ohm's law for the helical equilibrium of the uniform profile simulation with $S = 3 \times 10^4$ and $P = 300$. Figs. 3a, 3b and 3c are contour plots on poloidal cross-section of the helical flux function $\psi$ (black contours) and $\eta \mathbf{J} \cdot \mathbf{B}/B^2$, $\mathbf{E}_0 \cdot \mathbf{B}/B^2$ and $-\nabla\phi \cdot \mathbf{B}/B^2$, respectively (color levels). Fig. 3d represents contributions to parallel Ohm's law half way the dashed $\psi$ contour: $\eta \mathbf{J} \cdot \mathbf{B}/B^2$ (black line), $\mathbf{E}_0 \cdot \mathbf{B}/B^2$ (blue line) and $\nabla\phi \cdot \mathbf{B}/B^2$ (red line).

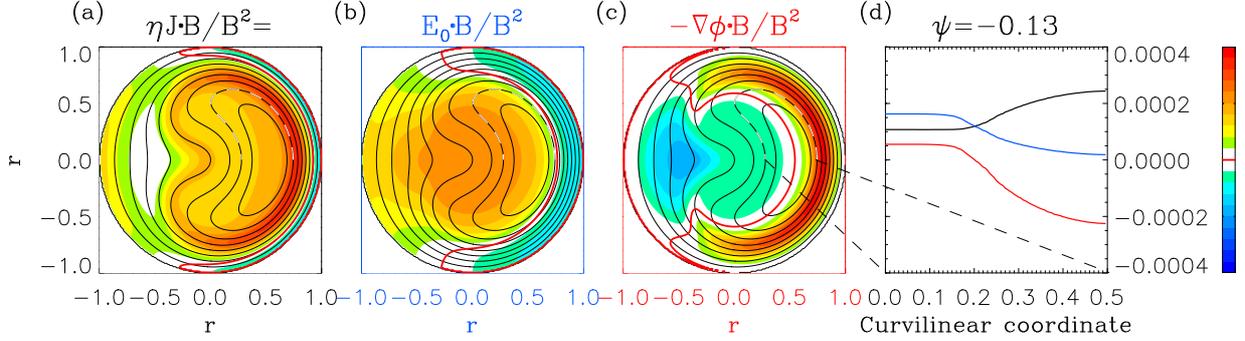

**FIGURE 4.** Contributions to parallel Ohm's law for the helical equilibrium of the standard profile simulation with $S = 3\times10^4$ and $P = 300$. Refer to the caption of Fig. 3 for an explanation on how to read the figure.

## IV. SUMMARY AND REMARKS

The impact of a uniform plasma resistivity in MHD modelling of helical RFP solutions has been considered. A flattening of the resistivity profile is found to cause a reduction of the dynamo action, which brings to a decrease of the toroidal field reversal. These findings are explained on the basis of the electrostatic nature of the RFP dynamo. In fact, a finite electrostatic field is required to balance parallel Ohm's law and the dynamo electric field merely corresponds to this electrostatic field[10,11]. We have shown here that with uniform plasma resistivity this balance is more easily achieved, while, in the case of standard profile simulations, the presence of resistivity gradients along magnetic field lines requires an additional electrostatic field, which provides a larger dynamo action and hence a deeper reversal.

Since the essential feature for dynamo reduction is the vanishing of resistivity gradients along magnetic surfaces, we may argue that future single-helicity simulations with non trivial self-consistent resistivity profiles would also result in a shallowing of the field reversal. This could explain the difficulty to find deeply-reversed RFP solutions of the helical Grad-Shafranov equation coupled with Ohm's law[13].

## ACKNOWLEDGMENTS

One of the authors (S. C.) wishes to thank J. F. Drake for his valuable suggestion on the effect of resistivity gradients along magnetic surfaces at the Festival de Theorie 2003 in Aix-en-Provence.